\documentstyle[preprint,aps,epsf]{revtex}

\def\mm{\mu^+\mu^-}

\def\mchi{m_{\tilde{\chi}^{\pm}}}
\def\msnu{m_{\tilde{\nu}}}
\begin{document}

\preprint{
\font\fortssbx=cmssbx10 scaled \magstep2
\hbox to \hsize{
\hfill$\vcenter{\hbox{\bf MADPH-98-1036}
             \hbox{\bf IUHET-381} \hbox{January
             1998}}$ } }

\title{\vspace*{.75in}
Chargino mass determination\\
at a muon collider}

\author{V. Barger$^a$, M.S.~Berger$^b$ and T.~Han$^{a}$}

\address{
$^a$Physics Department, University of Wisconsin, Madison, WI 53706, USA\\
$^b$Physics Department, Indiana University, Bloomington, IN 47405, USA}

\maketitle

\thispagestyle{empty}

\begin{abstract}

We analyze the prospects at
a muon collider for measuring chargino masses in the 
$\mu^+\mu^-\to \tilde{\chi}^+\tilde{\chi}^-$ process
in the threshold region. We find that for the lighter
chargino of a mass $100-200$~GeV, a measurement better than 
$50-300$~MeV should be possible with 50~fb$^{-1}$ integrated 
luminosity. The accuracy obtained here is better than with 
other techniques or at other facilities. The muon sneutrino 
mass, which enters through the $\tilde{\nu}_\mu$ exchange 
diagram, can also be simultaneously measured 
to a few GeV if it is not too heavy.
\end{abstract}

\newpage

\section{Introduction}

Particle masses can be measured to high precision through
threshold production cross sections at lepton colliders. 
This has been demonstrated at LEP II in $W$ pair production
at $\sqrt{s}=161$~GeV, just above $2M_W$. We have recently shown that future
high-luminosity $e^+e^-$ and $\mu^+\mu^-$ colliders can measure the $W$ boson,
top quark and Higgs boson masses at high precisions in the processes
$\ell^+\ell^-\to W^+W^-, t\overline{t}, ZH$\cite{wwthresh,zhthresh}.
Initial state radiation from muons is reduced compared to electrons, and 
muon colliders have negligible beamstrahlung which increasingly becomes a 
problem at linear electron colliders as the energy increases. Muon colliders 
thus could be very useful in precision measurements of particle
masses, widths, and couplings\cite{mupmumi,saus,montauk,sanfran,feas}.

In this paper we study the achievable accuracy in 
measuring the mass of the 
lighter chargino in the minimal supersymmetric
standard model (MSSM) via the cross section for 
\begin{equation}
\mm \to \chi^+ \chi^-
\end{equation}
near the threshold. The measurement
of the chargino mass via the threshold cross section has been considered 
previously for $e^+e^-$ machines in Ref.~\cite{oldthresh,nlcstudy}.
The narrower energy spread and the negligible beamstrahlung at a muon 
collider offer a distinct advantage over most electron-positron designs. 
We assume in this paper that 
the muon collider has a relatively modest (rms) beam energy spread of 
$R=0.1\%$. We consider 
a measurement with high integrated luminosity (50 fb$^{-1}$), 
carefully taking into account beam smearing effects and 
optimization of cuts to eliminate the background in the threshold region. 

A precision measurement of the chargino mass is a highly desirable goal 
to test
patterns of supersymmetry breaking. For example the relationship between
the lightest neutralino and the lighter chargino masses can be used to test
the existence of a universal soft SUSY-breaking parameter. Renormalization
Group Evolution (RGE) from the grand unification scale leads to the approximate
prediction $\mchi \simeq m_{{\tilde\chi}_2^0} \simeq 2m_{{\tilde\chi}_1^0}$ 
\cite{martin}. The predictions for chargino pair production 
have recently been investigated beyond the 
tree-level \cite{rc}. A precision measurement of the cross section
can test radiative corrections coming from heavy squarks, since the 
corrections depend on $\log(M_{\tilde{Q}}/m_{\tilde{\ell}})$.

The cross section of the chargino pair production
depends not only on $\mchi$ but also on the mass of the
muon sneutrino ($\msnu$) which enters through a $t$-channel
diagram. As we show in Sec.~II, a simultaneous measurement 
of both $\mchi$ and $\msnu$ is possible. In Sec.~III, we
compare our results with that achievable at an $e^+e^-$ linear
collider and with the kinematical end-point technique. 
We also comment on 
the benefits with polarized muon beams in studying the 
chargino mass and properties.

\section{Achievable Accuracy in $\mchi$}

If the lighter chargino is gaugino-dominated as expected \cite{rge,rge2}, 
then changing the parameters of the 
chargino mass matrix essentially changes the mass but 
does not significantly change its couplings.
 The chargino mass matrix is 
\begin{equation}
M_C=\left( \begin{array}{c@{\quad}c}
M_2 & \sqrt{2}M_W\sin \beta \\
\sqrt{2}M_W\cos \beta & -\mu 
\end{array} \right)\;,
\end{equation}
and in supergravity models the diagonal terms are expected to be larger
than the off-diagonal ones. As a typical illustration, we choose
the representative MSSM parameters
\begin{equation}
M_2=120\, {\rm GeV}, \quad \mu = 400\, {\rm GeV},\quad \tan\beta=4,
\label{para}
\end{equation}
where $M_2$ is the gaugino mass parameter, $\mu$ the
Higgs mixing and 
$\tan\beta=v_2/v_1$ the ratio of the vevs
of the two Higgs doublets in the MSSM.
The choice of Eq.~(\ref{para}) is motivated by the
``gaugino point'' of Ref.~\cite{fs}, so that the lighter
chargino is gaugino-like ($M_2<|\mu|$). This choice
corresponds to $\mchi=123$~GeV.

For the chargino pair production under discussion, the
sneutrino contribution in the $t$-channel
interferes destructively with the $s$-channel graphs.
Therefore one can envision a measurement of the 
cross section that essentially depends on just two parameters, 
$m_{\tilde{\chi}^{\pm}}$ and $m_{\tilde{\nu}}$.
Figure 1 illustrates the
total cross sections versus the center-of-mass
energy near threshold for various values of sneutrino 
mass, with other parameters as in Eq.~(\ref{para}).
The rapid rise of the cross section near threshold
is due to the $S$-wave pair production 
of spin-1/2 particles with small decay widths.
The cross section is typically of order 1 pb. Thus a
large signal sample of order $5\times 10^4$ chargino
events would be obtained with the assumed collider luminosity.

A simultaneous measurement of the chargino and sneutrino masses requires a 
sampling of the cross section at least two points. As in other threshold
measurements, the statistical precision on the chargino mass is maximized 
at CM energy $\sqrt{s}$ 
just above $2m_{\tilde{\chi}^{\pm}}$. However as is evident from Fig.~1, 
a change in the cross section at $\sqrt{s}=2m_{\tilde{\chi}^{\pm}}+1$~GeV
can also be due to a variation in the sneutrino mass, 
so a second measurement of 
the cross section must be made at a higher $\sqrt{s}$ where the dependence
of the cross section on the chargino mass and the slepton mass is different.
It turns out to be advantageous for the chargino mass measurement to choose 
this higher energy measurement at a $\sqrt{s}$ 
point where the chargino cross section 
is not flat. The precision that can be obtained in the chargino
mass depends substantially on the chargino mass itself since the cross-section
decreases with increasing $\mchi$. 
The heavier the chargino is, the less accurate the measurement
for a given luminosity.

The chargino decay mode is 
$\tilde{\chi}^{\pm}\to \tilde{\chi}^0f\overline{f}^\prime$,
resulting in large missing energy due to $\tilde{\chi}^0$ in the
final state, which is stable in the MSSM and thus escapes the 
detector. If $m_{\tilde{\chi}^{\pm}}-m_{\tilde{\chi}^0}>M_W$ 
then real $W$ contributions (two-body decay) dominate
and the $\chi^+ \chi^-$ 
final state is comprised of 49\% purely hadronic events,
42\% mixed hadronic-leptonic events, and 9\% purely leptonic events 
(these ratios are determined by the $W$ branching fractions). 
To effectively suppress the backgrounds, we concentrate
on the pure hadronic channel. 
The width of the chargino, typically less than a few MeV,
has a negligible impact on the threshold cross section 
even for the two-body decay case, provided that the lighter chargino
is gaugino-dominated. Based on the cross sections given
in Fig.~1 and including the decay branching ratios and signal efficiencies, 
the signal rate at $\sqrt{s} =2\mchi +1$~GeV would be about 20 fb 
for most values of the sneutrino mass. 
With 50 fb$^{-1}$ integrated luminosity, the cross section could be
measured to a statistical accuracy of about 3\%. 
Thus an understanding of the background to 
at least this level is necessary.

There are several backgrounds to the chargino pair signal, 
by far the largest being $\mu^+\mu^-\to W^+W^-$. 
The backgrounds have been studied in 
Refs.~\cite{tfmyo,grivaz}, and signal efficiencies were obtained 
for the various final states when the center-of-mass energy is 
$\sqrt{s}=500$~GeV. The dominant $W^+W^-$ background 
can be effectively eliminated by angular cuts because the $W$'s are produced 
in the very-forward direction. However, if the energy is reduced for 
running in the chargino threshold region, then the effectiveness
of the angular cuts would be reduced since the background events become
more spherical. Therefore we reinvestigate the acceptance criteria
near the threshold.

\begin{center}
\epsfxsize=4.75in\hspace{0in}\epsffile{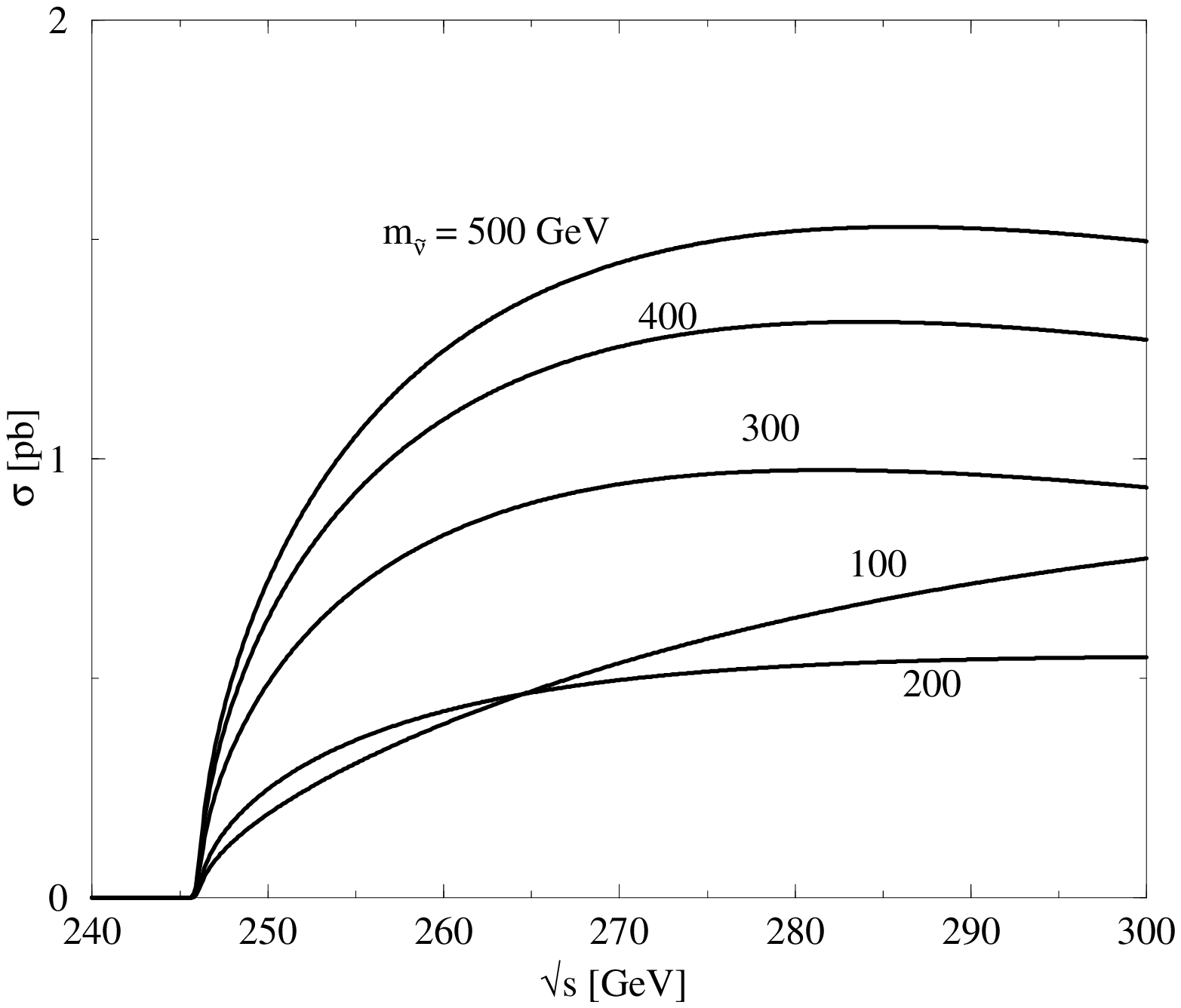}
\end{center}

\medskip
{\footnotesize Fig.~1: The cross section for 
$\mu^+\mu^-\to \tilde{\chi}^+\tilde{\chi}^-$ in the threshold region
for various sneutrino masses, with the parameters in Eq.~(\ref{para}). 
The sneutrino mass dependence arises from a 
$t$-channel contribution which interferes destructively with the 
$s$-channel diagrams. The muon collider is 
assumed to have a beam energy spread of $R=0.1\%$, 
and initial state radiation is included.}

Based on the characteristic kinematics of the signal,
we impose the following cuts to remove the backgrounds,
mainly from $W^+W^- \to 4$~jets:
\begin{itemize}
\item A cut on missing mass, roughly 
$2M_{{\tilde \chi}_0} < M({\rm miss}) <  2M_{{\tilde \chi}_0}+20$~GeV.

\item Require $\cos(\theta_{\rm W-miss}) > - 0.8$ where 
$\cos(\theta_{\rm W-miss})$ is the minimum cosine of the angle 
between the reconstructed faster $W$ and the missing momentum.

\item Require the reconstructed $W$'s be in the central region:
$|\cos(\theta _W)|<0.7$.

\end{itemize}
These cuts greatly reduce the $WW$ background to a negligible level.
The overall signal efficiency with these cuts is about 10\% for 
the fully hadronic decays.

\begin{center}
\epsfxsize=4.75in\hspace{0in}\epsffile{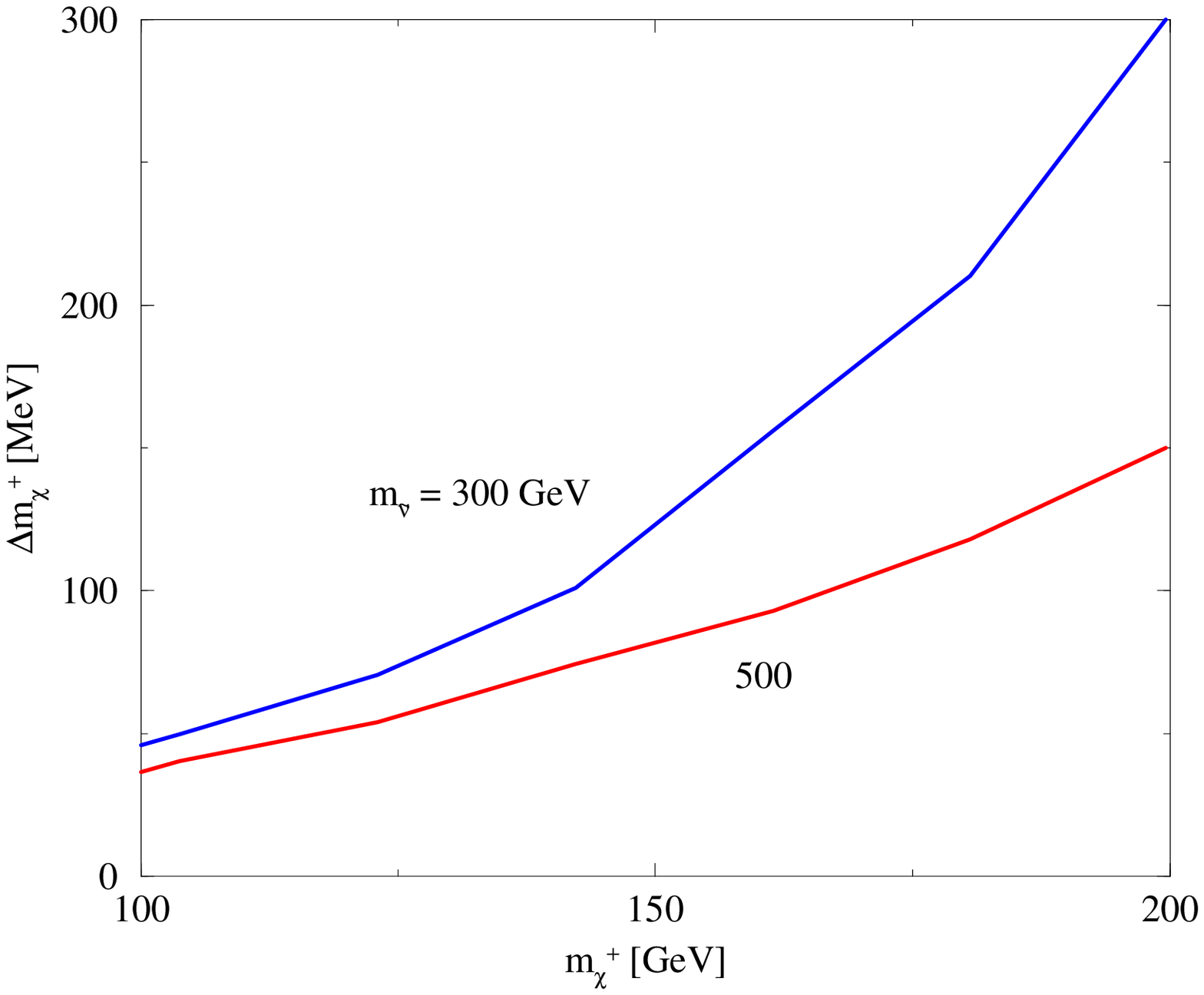}
\end{center}

\medskip
{\footnotesize Fig.~2: The $1\sigma$ precision obtainable in the 
chargino mass taking $m_{\tilde{\nu}} = 300$ and $500$~GeV assuming 
50~fb$^{-1}$ integrated luminosity. The precision on $\mchi$
is better for {\it larger} sneutrino mass (see Fig. 1).}

Figure~2 shows the expected precision of $\mchi$ from fully hadronic
decays with 50~fb$^{-1}$ integrated luminosity 
and a sneutrino mass of 300 and 500~GeV. 
For a lighter sneutrino, for which the destructive interference between
the $s$-channel and $t$-channel graphs is more severe, 
the precision of $\mchi$ is less. 
In the range of $\mchi=100-200$~GeV, a measurement
better than $50-300$~MeV is possible, much below the 1\% level.
The precision decreases with increasing chargino mass
since the production cross section decreases.

\begin{center}
\epsfxsize=4.75in\hspace{0in}\epsffile{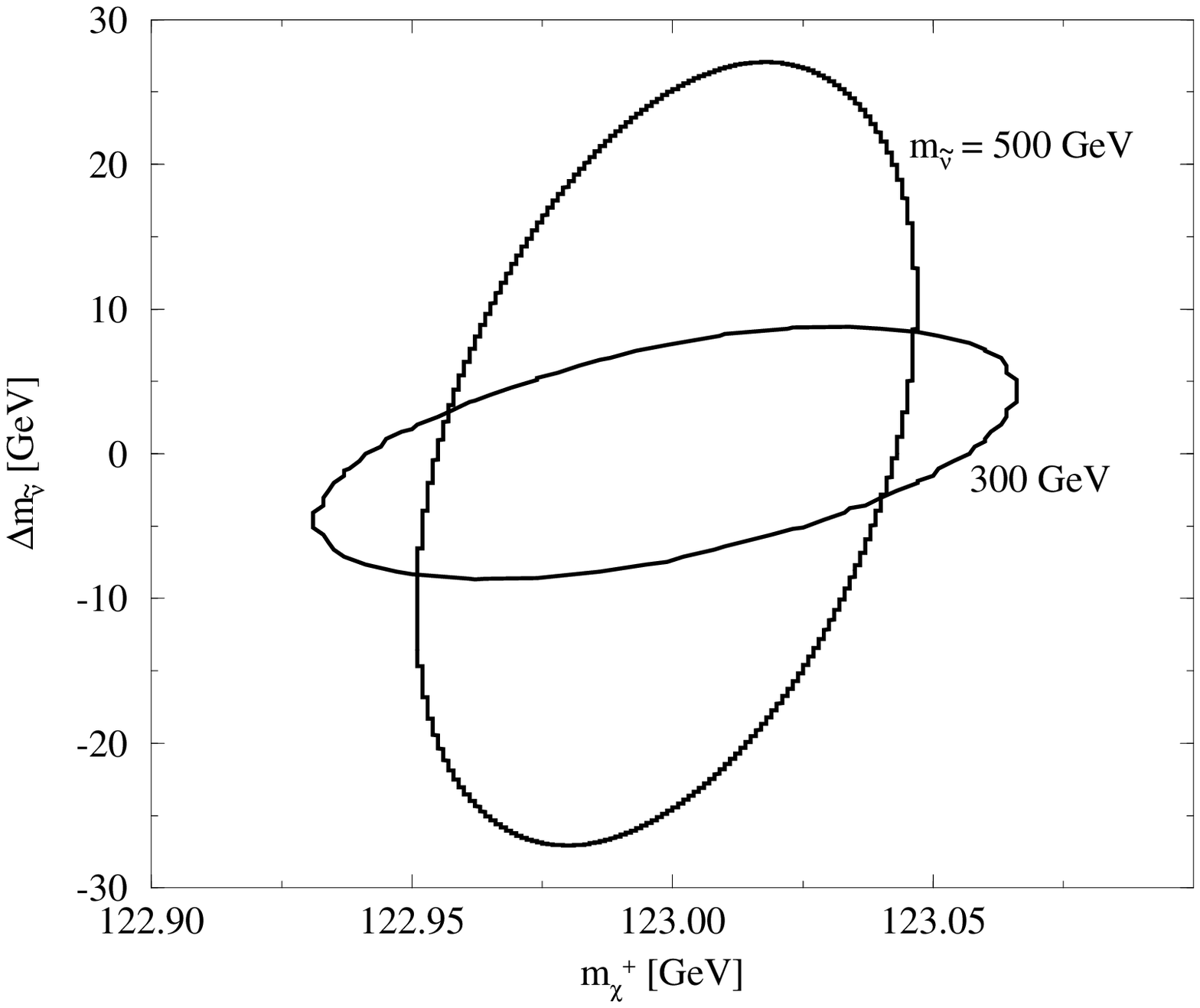}
\end{center}

\medskip
{\footnotesize Fig.~3: The $\Delta \chi^2 =1$ contours in the chargino 
mass - sneutrino mass plane, taking the parameters in Eq.~(\ref{para}) 
and $m_{\tilde{\nu}}=300$ and $500$~GeV. 
The curves assume 25~fb$^{-1}$ of integrated
luminosity is devoted to $\sqrt{s}=2m_{\tilde{\chi}^{\pm}}+1$~GeV, and 
25~fb$^{-1}$ is applied at $\sqrt{s}=2m_{\tilde{\chi}^{\pm}}+20$~GeV.}

The result of a fit to the chargino event rate is shown in Fig.~3, 
taking the parameters in Eq.~(\ref{para}) and assuming an integrated 
luminosity of 50~fb$^{-1}$. 
The cross section is measured just above the threshold 
$\sqrt{s}=2m_{\tilde{\chi}^{\pm}}+1$~GeV, and at a point well
above the threshold, $\sqrt{s}=2m_{\tilde{\chi}^{\pm}}+20$~GeV
(with 25 fb$^{-1}$ at each measurement). The chargino mass 
determination is better for higher sneutrino mass.
The cross section is more sensitive to $m_{\tilde{\nu}}$
when it is lighter, resulting in a better measurement of the 
sneutrino mass.
The sneutrino mass can be measured to about 6~GeV accuracy for 
$m_{\tilde{\nu}} = 300$~GeV and to about 20~GeV accuracy for 
$m_{\tilde{\nu}} = 500$~GeV. This provides an indirect method of 
measuring the sneutrino mass \cite{fs}, 
which would be especially valuable 
when the threshold for sneutrino pair production is not open.

\section{Discussions}

Comparing our results with similar studies for $e^+e^-$ colliders,
we find that the beam energy spread can cause a significant reduction
in the precision of the threshold measurement.
The most recent TESLA design envisions 
an electron beam energy spread of $R=0.2\%$\cite{miller} while the
Next Linear $e^+e^-$ Collider (NLC) 
design anticipates a beam energy spread of $R=1.0\%$. The NLC will
be able to achieve precisions which are from 15\% to 90\%
worse (for $\mchi$ from 100~GeV to 200~GeV)
than for the muon collider considered here, while the TESLA design should
achieve precisions less than 10\% worse than the muon collider. A high 
energy $e^+e^-$ collider in a Very Large Hadron Collider (VLHC)
tunnel would have a beam spread of $\sigma _E=0.26$~GeV\cite{norem} 
and would obtain results
with a precision comparable to those considered here.

The mass of the chargino can also be measured by finding the endpoint in 
the spectrum (or by fitting to the full spectrum) 
of the chargino decay products\cite{nlcstudy,tfmyo,hpssy,lykken,paige}. 
The endpoint
is determined strictly by the kinematics of the decay 
$\tilde{\chi}^{\pm}\to \tilde{\chi}^0f\overline{f}^\prime$, so it
is sensitive to both the chargino and neutralino masses. However the expected 
precision of the end point method with 50~fb$^{-1}$ of integrated luminosity 
is greater than 1.5\%.

A further advantage of the threshold measurement is that the chargino mass
measurement is largely independent of how it subsequently decays. 
Distributions in
the final state observables, say e.g. $E_{jj}$ from the decay
$\tilde{\chi}^{\pm}\to \tilde{\chi}^0jj$\cite{tfmyo}, depend on the 
neutralino mass. The cross section for chargino pair production, on the 
other hand, is independent of the final state decays, and only the 
branching fractions and detector efficiencies for the various final states
impact this measurement (if $m_{\tilde{\chi}^{\pm}}-m_{\tilde{\chi}^0}>M_W$
the branching fractions of chargino decay is given essentially in terms of 
the $W$ branching fractions).

We have assumed here that the chargino is lighter than the 
muon sneutrino, as is normally the case in mSUGRA models \cite{rge,rge2}.
If that is not so, the chargino has a new decay mode:
$\tilde{\chi}^{\pm}\to \ell^{\pm}\tilde{\nu}$. The signal efficiency 
of the cuts against background would need to be reconsidered if 
this mode is kinematically allowed.

It is expected that the both beams of a muon collider can be partially 
polarized, although with some loss of luminosity for high
polarization \cite{feas}. Polarization could prove a useful 
tool for studying the chargino pair production. When the chargino 
is gaugino-dominated, it couples to the left-handed $\mu^-$
because it is then dominantly the partner to the $W$. 
Should the chargino be Higgsino-dominated, one would 
want to employ right-handed $\mu^-$ polarization
since the $W^+W^-$ background would then be largely reduced.
In addition, the $t$-channel sneutrino exchange contribution
can be turned off by operating with a right-handed polarized 
$\mu^-$ beam. 

For the gaugino-dominated chargino considered here, both the signal and 
background are approximately 
proportional to $(1-P)^2$ where $P\equiv P_{\mu^-}=-P_{\mu^+}$ 
is the polarization of the two muon beams ($P=-1$ for a pure
left-handed ${\mu^-}$ or a right-handed ${\mu^+}$). 
The background ($W$ pairs) and the $t$-channel sneutrino
signal contribution couple to the left-handed 
$\mu ^-$ (and right-handed $\mu ^+$) beam. In the limit of 
$SU(2)\times U(1)$ symmetry, the $U(1)$ gauge boson couples only to the 
Higgsino component of the lighter chargino \cite{fs,mp}. 
So the $s$-channel graph also couples predominantly to 
the left-handed $\mu^-$ when the lighter chargino is gaugino-like as 
considered here. Thus for 100\%
polarized $\mu ^+$ and $\mu^- $ beams
the mass determination would improve by a factor of two
assuming the same integrated luminosity.

We have assumed in this study that the overall normalization of 
the chargino cross section is theoretically known,
apart from the contribution from the $t$-channel diagram
from a sneutrino of unknown mass. 
One can relax this assumption and allow the cross section
normalization to be another free parameter. 
Then at least three measurements for the cross section 
would be required to extract the two masses
($\mchi,m_{\tilde{\nu}}$) and the cross section normalization.
This would test the theoretical prediction for radiative corrections 
from which the mass scale of squarks might be inferred \cite{rc}. 
On the other hand, if the sneutrino is discovered independently 
and its mass reasonably well measured, one could carry out the 
two-point measurement, as presented this paper, to determine 
$\mchi$ and the cross section normalization.

\section{Conclusions}
A measurement of the lighter chargino mass to 
better than  $50-300$~MeV is possible for $\mchi=100-200$~GeV
by measuring the pair production cross section near 
threshold at a muon collider with 50~fb$^{-1}$ luminosity. 
This is superior to other techniques, such as the kinematical
end-point method, or at other colliders.
Only modest beam energy resolution ($R\sim 0.1\%$) 
is needed for the threshold
measurements.
The muon sneutrino mass can also be simultaneously measured 
to a few GeV if it is not too heavy.

\section*{Acknowledgments}

We thank J.~F.~Gunion for helpful
discussions at the initial stages of this
work. This work was supported in part by the U.S. Department of Energy
under Grants No. DE-FG02-95ER40896 and
No.~DE-FG02-91ER40661.
Further support was provided
by the University of Wisconsin Research
Committee, with funds granted by the Wisconsin Alumni Research
Foundation.

\end{document}